\documentclass[twocolumn,trackchanges]{aastex631}
\usepackage{newtxtext,newtxmath}

\newcommand{\mosel}{{\tt MOSEL}}

\newcommand{\zfourge}{{\tt ZFOURGE}}
\newcommand{\jades}{{\tt JADES}}

\newcommand{\oiii}{[\hbox{{\rm O}\kern 0.1em{\sc iii}}]\,5007}
\newcommand{\nii}{[\hbox{{\rm N}\kern 0.1em{\sc ii}}]\,6583}
\newcommand{\sii}{[\hbox{{\rm S}\kern 0.1em{\sc ii}}]\,6717,6731}
\newcommand{\siii}{[\hbox{{\rm S}\kern 0.1em{\sc iii}}]\,9069,9531}

\newcommand{\ciii}{[\hbox{{\rm C}\kern 0.1em{\sc iii}}]\,1907,1909}
\newcommand{\civ}{\hbox{{\rm C}\kern 0.1em{\sc iv}}\,1550}
\newcommand{\oiiiuv}{[\hbox{{\rm O}\kern 0.1em{\sc iii}}]\,1660,1666}
\newcommand{\oiiite}{[\hbox{{\rm O}\kern 0.1em{\sc iii}}]\,4363}
\newcommand{\heii}{\hbox{{\rm He}\kern 0.1em{\sc ii}}\,1640}
\newcommand{\hii}{\hbox{{\rm H}\kern 0.1em{\sc ii}}}

\newcommand{\oii}{[\hbox{{\rm O}\kern 0.1em{\sc ii}}]\,3727}
\newcommand{\hb}{\hbox{{\rm H}\kern 0.1em{\sc $\beta$}}}
\newcommand{\halpha}{\hbox{{\rm H}\kern 0.1em{\sc $\alpha$}}}

\newcommand{\ciiradio}{[\hbox{{\rm C}\kern 0.1em{\sc ii}}]$_{\rm 158 \mu m}$}

\newcommand{\SiII}{[\hbox{{\rm Si}\kern 0.1em{\sc ii}}]}

\newcommand{\OI}{[\hbox{{\rm O}\kern 0.1em{\sc i}}]}

\newcommand{\oiiiradio}{[\hbox{{\rm O}\kern 0.1em{\sc iii}}]$_{\rm 88 \mu m}$}

\newcommand{\msun}{${\rm M_{\odot}}$}

\newcommand{\lya}{\hbox{{\rm Lyman-}\kern 0.1em{\sc $\alpha$}}}

\newcommand{\SiIIAbsA}{\hbox{{\rm Si}\kern 0.1em{\sc ii}$\,\lambda1190$}}
\newcommand{\SiIIAbsB}{\hbox{{\rm Si}\kern 0.1em{\sc ii}$\,\lambda1260$}}
\newcommand{\SiIIAbsC}{\hbox{{\rm Si}\kern 0.1em{\sc ii}$\,\lambda1526$}}
\newcommand{\CIIAbs}{\hbox{{\rm C}\kern 0.1em{\sc ii}$\,\lambda1334$}}

\newcommand{\SiIVAbs}{\hbox{{\rm Si}\kern 0.1em{\sc iv}$\,\lambda\lambda1393,1402$}}
\newcommand{\CIVAbs}{\hbox{{\rm C}\kern 0.1em{\sc iv}$\,\lambda\lambda1548,1550$}}

\newcommand{\logmstar}{$\log(M_*/{\rm M}_{\odot})$}

\newcommand{\fesc}{$f_{esc}$}

\begin{document}

\title{MOSEL survey: Spatially offset Lyman-continuum emission in a new emitter at z=3.088}

\author[0000-0002-8984-3666]{Anshu Gupta}
\affiliation{International Centre for Radio Astronomy Research (ICRAR), Curtin University, Bentley WA, Australia}
\affiliation{ARC Centre of Excellence for All Sky Astrophysics in 3 Dimensions (ASTRO 3D), Australia}

\author[0000-0001-6324-1766]{Cathryn M. Trott}
\affiliation{International Centre for Radio Astronomy Research (ICRAR), Curtin University, Bentley WA, Australia}
\affiliation{ARC Centre of Excellence for All Sky Astrophysics in 3 Dimensions (ASTRO 3D), Australia}

\author[0000-0003-2035-3850]{Ravi Jaiswar}
\affiliation{International Centre for Radio Astronomy Research (ICRAR), Curtin University, Bentley WA, Australia}
\affiliation{ARC Centre of Excellence for All Sky Astrophysics in 3 Dimensions (ASTRO 3D), Australia}

\author{E. V. Ryan-Weber}
\affiliation{Centre for Astrophysics and Supercomputing, Swinburne University of Technology, Hawthorn, VIC 3122, Australia}
\affiliation{ARC Centre of Excellence for All Sky Astrophysics in 3 Dimensions (ASTRO 3D), Australia}

\author{Andrew J. Bunker}
\affiliation{Department of Physics, University of Oxford, Denys Wilkinson Building, Keble Road, Oxford OX1 3RH, UK}

\author[0000-0003-4804-7142]{Ayan Acharyya}
\affiliation{Department of Physics and Astronomy, Johns Hopkins University, Baltimore, MD 21218, USA}

\author{Alex J. Cameron}
\affiliation{Department of Physics, University of Oxford, Denys Wilkinson Building, Keble Road, Oxford OX1 3RH, UK}

\author[0000-0001-6003-0541]{Ben Forrest}
\affiliation{Department of Physics and Astronomy, University of California Davis, One Shields Avenue, Davis, CA, 95616, USA}

\author[0000-0003-1362-9302]{Glenn G. Kacprzak}
\affiliation{Swinburne University of Technology, Hawthorn, VIC 3122, Australia}
\affiliation{ARC Centre of Excellence for All Sky Astrophysics in 3 Dimensions (ASTRO 3D), Australia}

\author[0000-0003-2804-0648]{Themiya Nanayakkara}
\affiliation{Centre for Astrophysics and Supercomputing, Swinburne University of Technology, Hawthorn, VIC 3122, Australia}
\affiliation{ARC Centre of Excellence for All Sky Astrophysics in 3 Dimensions (ASTRO 3D), Australia}

\author[0000-0001-9208-2143]{Kim-Vy Tran}
\affiliation{School of Physics, University of New South Wales, Sydney, NSW 2052, Australia}
\affiliation{ARC Centre of Excellence for All Sky Astrophysics in 3 Dimensions (ASTRO 3D), Australia}
\affiliation{Center for Astrophysics $|$ Harvard \& Smithsonian, Cambridge, MA }

\author[0000-0003-1130-6390]{Aman Chokshi}
\affiliation{The University of Melbourne, School of Physics, Parkville, VIC 3010, Australia}
\affiliation{ARC Centre of Excellence for All Sky Astrophysics in 3 Dimensions (ASTRO 3D), Australia}

\begin{abstract}
We present the discovery of a unique Lyman-continuum (LyC) emitter at $z=3.088$. The LyC emission were detected using the Hubble Space Telescope (HST) WFC3/UVIS F336W filter, covering a rest-frame wavelength range of $760-900\, \Angstrom$. The peak signal-to-noise ratio (SNR) of LyC emission is 3.9 in a $r=0.24''$ aperture and is spatially offset by $0.29''\pm0.04$ ($\sim 2.2\pm0.3$ kpc) from the rest-UV emission peak (F606W). By combining imaging and spectroscopic data from the James Webb Space Telescope (JWST) JADES, FRESCO and JEMS surveys, along with VLT/MUSE data from the MXDF survey, we estimate that the probability of random alignment with an interloper galaxy causing the LyC emission is less than $6\times 10^{-5}$. The interstellar medium (ISM) conditions in the galaxy are similar to other LyC emitters at high redshift ($12+\log(O/H)=7.79_{-0.05}^{+0.06}$, $\log U =-3.27_{-0.12}^{+0.14}$, $O32 = 3.65\pm0.22$), although the single-peaked \lya\ profile and lack of rest-UV emission lines suggest an optically thick ISM. We think that LyC photons are leaking through a narrow cone of optically thin neutral ISM, most likely created by a past merger (as evidenced by medium-band F210M and F182M images). Using the escape fraction constraints from individual leakers and a simple model, we estimate that the opening half-angle of ionization cones can be as low as $16^{\circ}$ (2\% ionised fraction) to reproduce some of the theoretical constraints on the average escape fraction for galaxies. The narrow opening angle required can explain the low number density of confirmed LyC leakers.
\end{abstract}

\keywords{galaxies: high-redshift – emission-line – interactions – evolution 
}


\section{Introduction} \label{sec:intro}

Understanding the escape of ionizing photons from the first galaxies remains one of the biggest challenges for the models of galaxy formation and evolution. The past year of operations by the James Webb Space Telescope (JWST) has significantly transformed our understanding of early-universe galaxies. Nevertheless, the high opacity of the intergalactic medium prevents us from detecting the ionizing photons escaping from the first galaxies. Therefore, we need to rely on analogs of first galaxies to understand the mode of Lyman continuum (LyC) escape.

In the past decade, photometric observations revealed elevated emission line equivalent widths (EW) for $z>6$ galaxies \citep{labbe_2013_SPECTRALENERGYDISTRIBUTIONS, roberts-borsani_2016_GALAXIESREDSPITZER, barro_2019_CANDELSSHARDSMultiwavelength, mainali_2019_RELICSSpectroscopyGravitationallylensed, endsley_2020_IiiEquivalentWidth}. Recent observations with JWST have confirmed this, demonstrating that 80-90\% of galaxies within the first billion years of the universe have \oiii+\hb\ EW $>800\,\Angstrom$, almost three times the EW of a typical star-forming galaxy at $z\sim2$ \citep{endsley_2023_StarformingIonizingProperties, cameron_2023_JADESProbingInterstellar, tang_2023_JWSTNIRSpecSpectroscopy, rinaldi_2023_MIDISStrongHv, boyett_2024_ExtremeEmissionLine}. Consequently, many observational programs focus on studying the detailed properties of extreme emission line galaxies (EELGs), where emission lines from gas contribute almost 40-50\% to the total flux in certain photometric bands, both at low \citep{yang_2017_LyaProfileDust, yuma_2019_GiantGreenPea, izotov_2018_LowredshiftLymanContinuum, lumbreras-calle_2022_JPLUSUncoveringLarge} and high redshifts \citep{atek_2011_VERYSTRONGEMISSIONLINE, maseda_2014_NatureExtremeEmission, gupta_2022_MOSELSurveyExtremely}. EELGs typically exhibit high specific star formation rates and high ionization parameters \citep{tang_2019_MMTMMIRSSpectroscopy, gupta_2022_MOSELSurveyExtremely} that could be linked with a high LyC escape fraction \citep{izotov_2018_LowredshiftLymanContinuum}.
 


Observations of dwarf star-forming galaxies (SFGs) in the nearby universe with HST Cosmic Origin Spectrograph have built a sizable population of local LyC leakers \citep{izotov_2016_DetectionHighLyman, izotov_2016_EightCentLeakage, izotov_2018_LowredshiftLymanContinuum, flury_2022_LowredshiftLymanContinuuma}. The escape fraction shows a weak positive correlation with \oiii/\oii\ ratio and an inverse correlation with the velocity separation between the blue and red peaks of the \lya\ emission \citep{izotov_2018_LowredshiftLymanContinuum, flury_2022_LowredshiftLymanContinuum}. These observations suggest that a harder ionizing field and/or a low neutral gas covering fraction are crucial to facilitate the leakage of LyC photons.

At $z\sim1-3$, the stacking experiments with SFGs yield an escape fraction between $f_{esc}=1\%-10\%$ \citep{steidel_2018_KeckLymanContinuum, wang_2023_LymanContinuumEscape}. The individual detection of LyC leakers at high redshifts is complicated by the higher chance of contamination from interloper galaxies \citep{vanzella_2010_GreatObservatoriesOrigins,vanzella_2012_DETECTIONIONIZINGRADIATION} and attenuation by the partially neutral intergalactic medium (IGM) \citep{inoue_2014_UpdatedAnalyticModel, bassett_2021_IGMTransmissionBias}. Deep imaging campaigns with HST and u-band imaging on ground-based telescopes have created significant samples of candidate LyC leakers at cosmic noon \citep[e.g.,][]{mestric_2020_OutsideLymanbreakBox, kerutt_2023_LymanContinuumLeaker}. Recently, \cite{wang_2023_LymanContinuumEscape} used deep UVIS imaging from the UVCANDELS survey to report a tentative discovery of five LyC leakers between redshift $2.4<z<3.7$, some of them showing $f_{esc}>60\%$. However, follow-up spectroscopy is required to rule out any possible foreground contaminant.

The A2218-Flanking and the Sunburst arc are two gravitationally lensed LyC leakers at $z\sim 2.5$. In A2218-Flanking, LyC radiation is leaking from a compact dwarf galaxy \citep{bian_2017_HighLymanContinuum}, whereas in the Sunburst Arc, LyC radiation is leaking from a compact star-forming region that could be a young massive star cluster \citep{rivera-thorsen_2019_GravitationalLensingReveals}. Ion1 \citep{vanzella_2012_DETECTIONIONIZINGRADIATION, ji_2020_HSTImagingIonizing}, Ion2 \citep{debarros_2016_ExtremeIIIEmitter, vanzella_2016_HIGHRESOLUTIONSPECTROSCOPYYOUNG, vanzella_2020_IonizingIntergalacticMedium}, Ion3 \citep{vanzella_2018_DirectLymanContinuum, vanzella_2020_IonizingIntergalacticMedium}, and Q1549-C25 \citep{shapley_2016_Q1549C25CleanSource} are a few examples of non-lensed LyC leakers, each showing a wide array of interstellar medium conditions and escape fractions. For some of these galaxies, escape fractions can be higher than 50\% \citep{vanzella_2016_HIGHRESOLUTIONSPECTROSCOPYYOUNG, debarros_2016_ExtremeIIIEmitter, saxena_2022_NoStrongDependencea, marques-chaves_2022_NoCorrelationLyman, rivera-thorsen_2022_BottomupSearchLymancontinuum}. Recently, \cite{kerutt_2023_LymanContinuumLeaker} discovered five new LyC leakers using the MUSE-Wide survey \citep{urrutia_2019_MUSEWideSurveySurvey} by searching for LyC emission within a small aperture ($0.35''$) spatially coincident with peak rest-UV emission in \lya\ emitters.

The \lya\ emission profile for high redshift LyC leakers exhibits a range, contrary to the classic double-peaked \lya\ emission observed in most LyC leakers at $z\sim 0$. A few LyC emitters show 3-4 peaks in \lya\ emission indicative of multiple kinematic components in the neutral interstellar medium (ISM) \citep{vanzella_2020_IonizingIntergalacticMedium}. \lya\ emission is only single-peaked in Q1549-C25 \citep{shapley_2016_Q1549C25CleanSource} and appears in absorption in Ion1 ($z=3.78$), indicative of a highly optically thick ISM \citep{ji_2020_HSTImagingIonizing}. For Ion1, the LyC emission is spatially offset by $0.''12\pm 0.''03$ from the peak UV emission suggesting that the ionizing photons might be leaking from a small part of the galaxy. \cite{kerutt_2023_LymanContinuumLeaker} find a huge scatter between the properties of \lya\ emission and escape fraction for high redshift LyC emitters.
  
This paper reports the discovery of a new LyC leaker at $z=3.088$ in the Chandra Deep Field South \citep[CDFS, ][]{giacconi_2002_ChandraDeepField} and analyzes its ISM conditions using the wealth of ancillary data from the JWST and MUSE spectrograph \citep{bunker_2023_JADESNIRSpecInitial, bacon_2023_MUSEHubbleUltra}. The paper is organized as follows. Section \ref{sec:data} describes the photometric and spectroscopic data analysed here. In section \ref{sec:sed} and \ref{sec:lyc}, we discuss the spectral energy distribution fitting, the LyC detection and escape fraction measurements. In sections \ref{sec:nirspec} and \ref{sec:muse}, we report properties of ISM derived from JWST/NIRSpec and MUSE spectrum respectively. We discuss our results in Section \ref{sec:discussion} and present a simple model to derive the opening angle of the ionization cone. 
The paper uses a flat $\Lambda$CDM cosmology with $H_0 = 70$ km s$^{-1}$ Mpc$^{-1}$, $\Omega_{\rm M} = 0.3$, and $\Omega_\Lambda = 0.7$.

\begin{figure*}
	\centering
	\tiny
	\includegraphics[scale=0.45, trim=0.0cm 0.0cm 0.0cm 0.0cm,clip=true]{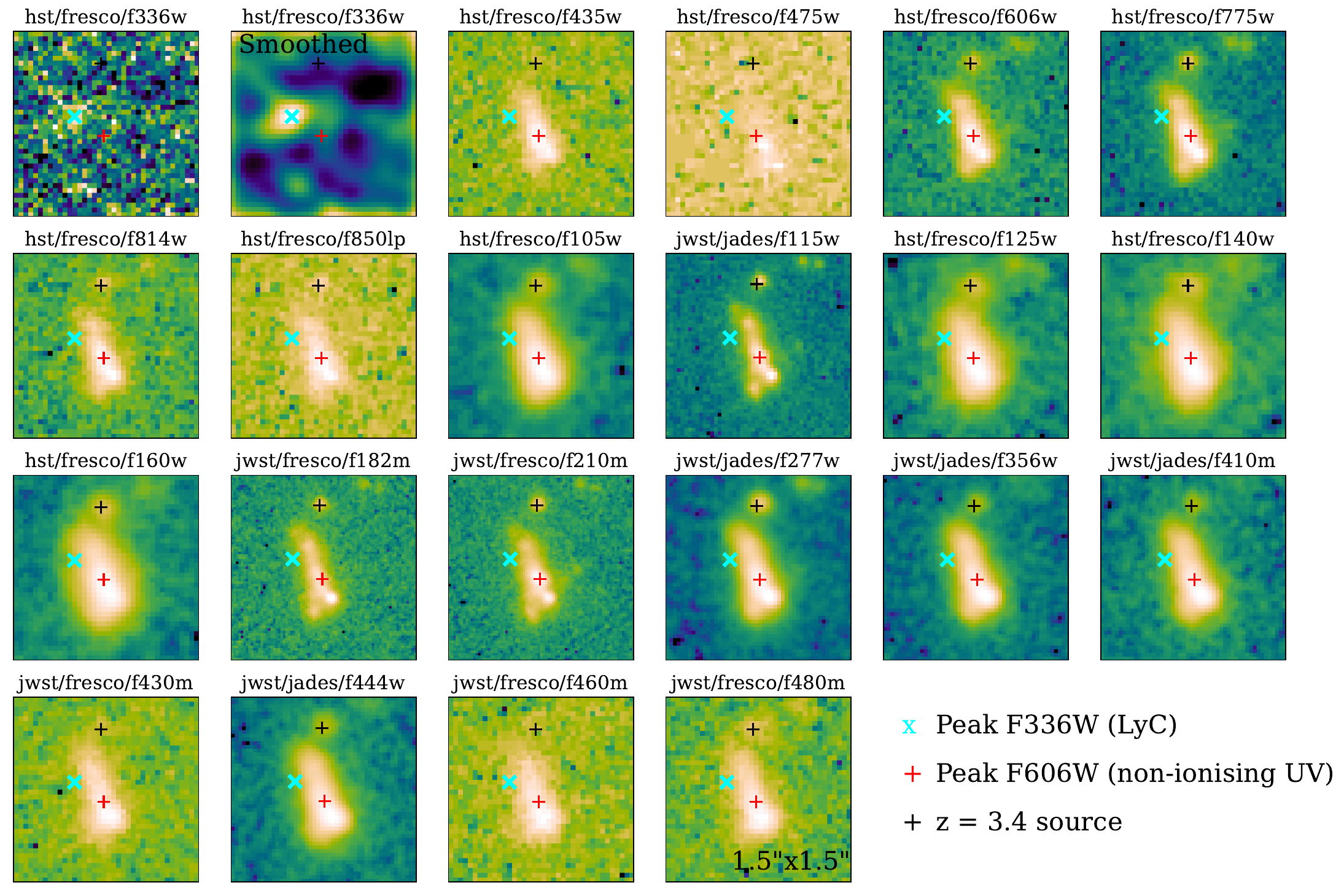}		
\caption{Multi-band imaging of galaxy z19863 from HST and JWST in increasing order of wavelength. Each panel is $1.5''\times 1.5''$ across. The top two panels correspond to the original and PSF-smoothed images from the HST F336W filter (LyC). The red plus and cyan cross mark the position of the peak non-ionising rest-UV (F606W) and the ionising rest-UV flux (F336W) respectively. The black plus marks the location of the $z=3.4$ \lya\ emitter detected in the MXDF survey (see Section \ref{sec:muse}) for more details. }
	\label{fig:fresco_imaging}
\end{figure*}

\section{Data}\label{sec:data}
The galaxy z19863 (RA=53.1699, DEC=-27.7684) was identified as an EELG by \cite{forrest_2018_ZFOURGEUsingComposite} using photometric data from the FourStar Galaxy Evolution survey \citep[\zfourge;][]{straatman_2016_FourStarGalaxyEvolution}. It is part of the Multi-Object Spectroscopy of Emission Line \citep[\mosel;][]{tran_2020_MOSELStrongOiii, gupta_2022_MOSELSurveyExtremely} survey of galaxies, with aims to analyse the properties of EELGs at $2.5<z<4$. The galaxy lies in the Hubble Ultra Deep Field (HUDF), providing us access to a multitude of ancillary data from both HST and JWST (Figure \ref{fig:fresco_imaging}).

  
We combine NIRCam imaging from the JWST Advanced Deep Extragalactic Survey 
\citep[\jades,][]{bunker_2023_JADESNIRSpecInitial, hainline_2023_CosmosItsInfancy, rieke_2023_JADESInitialData, eisenstein_2023_OverviewJWSTAdvanceda}, the FRESCO survey  \citep{oesch_2023_JWSTFRESCOSurvey}, and the JWST Extragalactic Medium-band Survey \citep[JEMS,][]{williams_2023_JEMSDeepMediumband}, giving us extensive broad and medium band coverage. Additionally, the FRESCO survey has released astrometrically aligned mosaics based on the archival photometry on the Hubble Legacy Fields including the UVIS/F336W imaging \citep{illingworth_2016_HubbleLegacyFields, whitaker_2019_HubbleLegacyField}.    

The galaxy has a spectroscopic redshift of $z=3.088$ determined from the deep NIRSpec spectrum from the \jades\
\citep{bunker_2023_JADESNIRSpecInitial}. We also use the data from the MUSE eXtremely Deep Field \citep[MXDF, ][]{bacon_2023_MUSEHubbleUltra} to rule out interloper galaxies and analyse the \lya\ emission profile. At $z=3.088$, the UVIS/F336W filter covers rest-frame wavelengths between $760-900\,\Angstrom$, and therefore should have no contamination from non-ionising photons.    


\section{Analysis}

\subsection{Spectral energy distribution}\label{sec:sed}
We convolve the images down to a common PSF of the NIRCam/F444W filter and use a $3\sigma$ threshold to create a segmentation map based on the F444W image. The fluxes are extracted within an aperture defined by KRON parameters of 2.0. A slightly lower KRON parameter is used here to minimize the contamination from a nearby background source at $z=3.4$. We estimate errors by taking the root mean square of flux at random locations within the same aperture for each image.

\begin{figure*}
	\centering
	\tiny
	\includegraphics[scale=0.4, trim=0.0cm 0.0cm 0.0cm 0.0cm,clip=true]{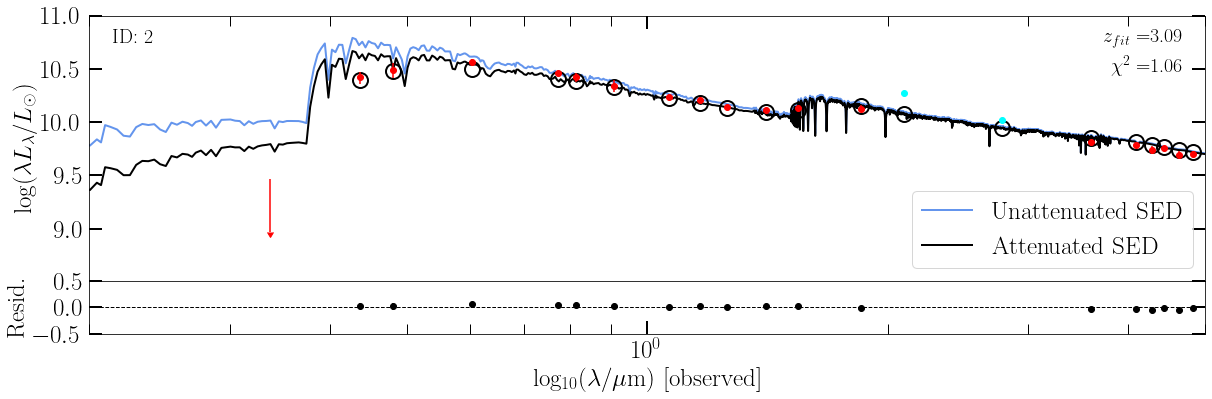}		
\caption{
The best-fit SED model using MAGPHYS before (blue curve) and after (black curve) correcting for dust attenuation. The red dots and open black circles correspond to the observed flux and derived flux from the best-fit SED respectively. The cyan dots represents the two filters (F210M and F277W) that are excluded while modelling the SED to minimise contamination by bright \oiii\ and \halpha\ emission lines. }
	\label{fig:SED}
\end{figure*}

We use the MAGPHYS \citep{dacunha_2008_SimpleModelInterpret, dacunha_2015_ALMASurveySubmillimeter} spectral energy distribution (SED) fitting code with the BC03 stellar population synthesis model \citep{bruzual_2003_StellarPopulationSynthesis}, delayed exponentially declining star formation history model, and \cite{charlot_2000_SimpleModelAbsorptiona} dust attenuation law to derive physical properties. We only use filters that are not contaminated by bright nebular emission lines such as \oiii\ and \halpha\ (all except F210M and F277W, Figure \ref{fig:SED}). We also do not include an IGM absorption model.
 
\subsection{LyC Detection}\label{sec:lyc}

We use the HST UVIS/F336W mosaics released by the FRESCO survey \citep{oesch_2023_JWSTFRESCOSurvey}, created using the Hubble Legacy Fields photometry \citep{teplitz_2013_UVUDFUltravioletImaging, rafelski_2015_UVUDFUltravioletNearinfrared, illingworth_2016_HubbleLegacyFields, whitaker_2019_HubbleLegacyField}, to visually search for LyC emission around \mosel\ targets. The mosaic relies on the archival HST imaging and hence has non-uniform coverage and depth across the FRESCO field of view (FOV). Out of 76 EELGs identified by \cite{forrest_2018_ZFOURGEUsingComposite} in the CDFS, 54 fall within the FRESCO FOV. We only found significant emission around one EELGs that happen to fall in the deepest region in the F336W image. Figure \ref{fig:fresco_imaging} shows the multi-band imaging for z19863, along with the unconvolved and PSF convolved F336W imaging, clearly indicating the possible LyC emission towards the NE side of the galaxy. 

\begin{figure}
	\centering
	\tiny
	\includegraphics[scale=0.45, trim=0.0cm 0.0cm 0.0cm 0.0cm,clip=true]{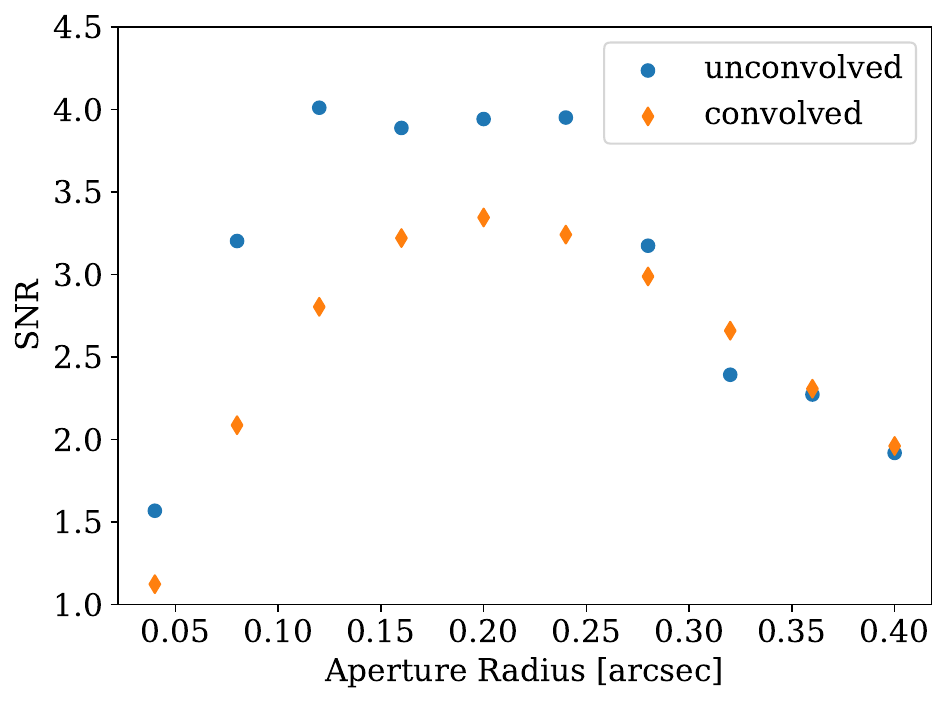}		
\caption{The SNR as a function of the aperture radius for the original (blue circle) and PSF-convolved (orange circle) F336W image. Each aperture is positioned at the peak LyC flux (cyan cross in Figure \ref{fig:fresco_imaging}). }
	\label{fig:snr_aperture}
\end{figure}

We estimate the signal-to-noise ratio (SNR) of LyC emission by placing an aperture of variable size at the pixel corresponding the peak emission in both the unconvolved and PSF-convolved F336W images (Figure \ref{fig:snr_aperture}). In the unconvolved image, we reach a maximum SNR of $4.0$ within an aperture of $r=0.''12$, corresponding to $m_{F336W}=28.7_{-0.2}^{+0.3}$. The SNR drops marginally to 3.95 at $r=0.''24$ corresponding to the $m_{F336W}=27.9_{0.2}^{+0.3}$, a slightly larger aperture size than the  FWHM ($0.''17$) of the F336W filter. The SNR drops to 2.4 at $r=0.32''$
and results in a marginal drop in flux ($m_{F336W}=28.0_{-0.3}^{+0.4}$). In the PSF-convolved image, the SNR reaches a maximum of 3.3 within an aperture of $r=0.2''$.  The PSF-convolved image does indicated some weak evidence of extended LyC emission. Using an elliptical aperture ($a=0.''28, b/a=0.7,\ \theta=120^{\circ}$), results in only a marginal increase in the SNR; therefore, significant flux is not lost by the choice of a circular aperture. 

    \begin{table}[t]
    \small
    \caption{Physical properties of z19863}
    \label{tb1}
            \begin{tabular}{p{3.cm} p{3.cm}}
            \hline
            \hline
            RA & 53.1699\\
            DEC & -27.7684\\
            z & 3.088 \\
            \logmstar\ & $9.2\pm 0.1$\\
            SFR$_{\halpha}$ [M$_{\odot}$/yr] & $4.3\pm0.09$\\
            SFR$_{SED}$ [M$_{\odot}$/yr] & $5.2\pm0.2$\\
            $E(B-V)_{gas}$ & $0.09\pm0.02$\\
            $12+log(O/H)$ & $7.79_{-0.05}^{+0.06}$\\
            $logU$ & $-3.27_{-0.12}^{+0.14}$\\
            \multicolumn{2}{c}{$L_{1500}/L_{900} = 6.66$, MAGPHYS}\\            $f_{esc}^{rel}$   & $2.8\pm0.7$\\
            $f_{esc}^{abs}$ & $1.2\pm0.3$\\
            \multicolumn{2}{c}{$L_{1500}/L_{900} = 1.25$, lower limit}\\
            $f_{esc}^{rel}$  & $0.5\pm0.1$\\
            $f_{esc}^{abs}$ & $0.24\pm0.06$\\
            
            \hline
        \end{tabular}
    \begin{flushleft}
    \end{flushleft}
    \end{table}

\begin{figure*}
	\centering
	\tiny
	\includegraphics[scale=0.65, trim=0.0cm 0.0cm 0.0cm 0.0cm,clip=true]{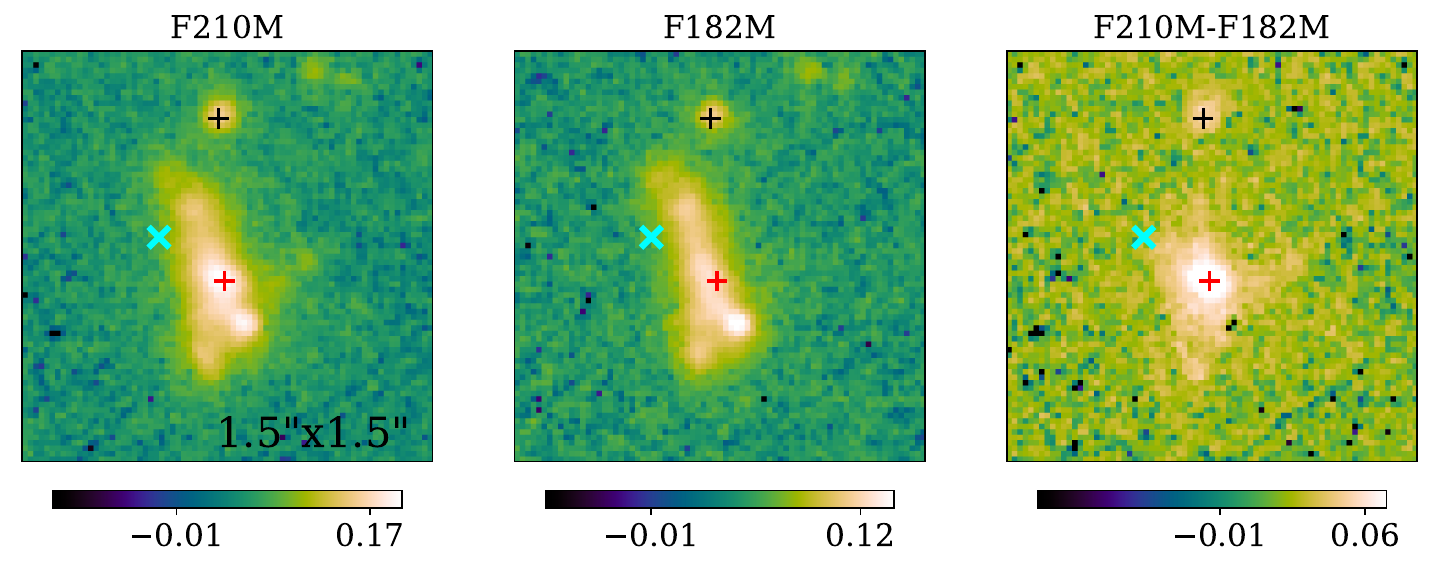}	
\caption{
JWST medium band F210M (left: stars+gas), F182M (middle: stars), and F210M-F182M (right: gas) images for the galaxy z19863 at $z=3.088$. The red plus marks the peak emission in F606W image (non-ionising UV emission), and the cyan cross marks the peak LyC emission in F336W.}
	\label{fig:fresco_pseudo_o3}
\end{figure*}

\subsubsection{Detection reliability test}

To test the reliability of our detection, we check for the growth curve  at random locations in the $1'\times1'$ region covering the deepest F336W data. We use the F160W segmentation map to mask out all bright targets in the F336W image. We then place two circular apertures of $r=0.12'',{\rm and}\ 0.24''$ at 20,000 random locations. We only measure the growth curve if the percentage of unmasked pixel within the largest aperture was greater than 90\%. Only one random location out of nearly 16470 possible locations  had S/N$\ge4.0$ within aperture $r=0.12''$ and S/N$\ge3.9$ within $r=0.24''$ aperture.  There is clear emission in the F606W image at the same random location, indicating that the false detection is an interloper galaxy. We do not detect any signature of any interlopers in either HST or JWST imaging and MUSE datacube coincident with the LyC detection. Thus the probability of randomly detecting a similar source is less than $p<6\times10^{-5}$.

\subsubsection{Spatial offset of ionising emission}

The peak non-ionizing UV emission (F606W) and ionizing emission have a spatial offset of $0.29''\pm0.04''$ or $2.2\pm0.3$ kpc at $z=3.088$. We checked for astrometric alignment between the F336W and F606W images by estimating the difference in the centroids of the 20 brightest sources in the deepest $1'\times1'$ FOV F336W image. The average difference in x-centroid [y-centroid] is $0\pm1$ [$0\pm1$] pixels, significantly smaller than the peak separation ($\sim 7$ pixels) between ionizing and non-ionizing emission from our target. Therefore, the astrometric inaccuracies cannot account for the spatial offset.

\subsubsection{LyC escape fraction}

The escape fraction is defined as the ratio of ionizing ($f_{\rm LyC}$ at $\lambda<912\,\Angstrom$) divided by the non-ionizing flux density \citep[$f_{\rm nLyC}$ typically at $1500\,\Angstrom$,][]{steidel_2001_LymanContinuumEmissionGalaxies}. We use the standard procedure to calculate escape fraction \citep[e.g.,][]{siana_2007_NewConstraintsLyman, ji_2020_HSTImagingIonizing}. As mentioned earlier, we use the F606W filter (rest-frame $\sim1500\,\Angstrom$) to measure non-ionizing photon flux density. The ionizing photons can sometimes escape from certain parts of the galaxy \citep{verhamme_2015_UsingLymanDetect}, therefore we use the same aperture to estimate flux for both ionizing and non-ionizing UV components. Using an aperture centered at the peak LyC flux and $r=0.12''$ gives $(f_{\rm LyC})_{\rm obs} = 12.5\pm 3.1$\,nJy and $(f_{\rm nLyC})_{\rm obs} = 11.7\pm1.2$\,nJy.
If we use an aperture of $r=0.24''$, then $(f_{\rm LyC})_{\rm obs} = 24.5\pm 6.2$\,nJy and $(f_{\rm nLyC})_{\rm obs} = 89.9\pm2.5$\,nJy. Thus, $(f_{\rm nLyC}/f_{\rm LyC})_{\rm obs} = 0.9 \pm 0.3$ within the $r=0.12''$ aperture and $(f_{\rm nLyC}/f_{\rm LyC})_{\rm obs} = 3.7 \pm 0.9$ within the $r=0.24''$ aperture. This is done after convolving the F336W and F606W images to a common PSF of F606W.

We need precise measurements of intrinsic non-ionizing to ionizing photon luminosity ($L_{1500}/L_{900}$), accurate IGM attenuation ($T_{\text{IGM}}$) at the redshift of our galaxy, and the dust attenuation ($A_{1500}$) to estimate relative and absolute escape fraction. We use the IGM transmission of $T_{\text{IGM}}=0.66$ at $z=3.08$ calculated by \cite{kerutt_2023_LymanContinuumLeaker}, which uses the mean transmission from the 5\% of the brightest sightlines based on the \cite{inoue_2014_UpdatedAnalyticModel} model. The adopted $T_{\text{IGM}}$ sits comfortably in the range calculated for $z=3.1$ \lya\ emitters by \cite[][see their Figure 6]{bassett_2021_IGMTransmissionBias}, taking into account the bias of detected LyC emitters along clean lines in the IGM.

Using $E(B-V)= 0.09\pm0.02$ (see section \ref{sec:dust}) and the extinction curve for the Small Magellanic Cloud (SMC) from \cite{gordon_2003_QuantitativeComparisonSmall}, we obtain $A_{1500} = 1.2\pm0.3$. We note that the NIRSpec spectra do not cover the region where LyC emission is detected. The region occupying the bulk of gas emitting \halpha\ and \hb\ emission lines can be different from the region with bright OB-type stars, as has been postulated by other studies \citep{vanzella_2020_IonizingIntergalacticMedium, vanzella_2022_HighStarCluster, mestric_2023_CluesPresenceSegregation}. Therefore, the extinction towards the LyC region could be significantly different.

Finally, the intrinsic luminosity ratio $L_{1500}/L_{900}$ can vary between $1.25-7$ depending on the assumptions in the stellar population model \citep{leitherer_1999_Starburst99SynthesisModels, eldridge_2017_BinaryPopulationSpectral}. The best-fit SED model from MAGPHYS (Section \ref{sec:sed}) gives $L_{1500}/L_{900} = 6.66$, resulting in $f_{\text{esc}}^{\text{rel}} = 2.8\pm 0.7$, and $f_{\text{esc}}^{\text{abs}} = 1.3\pm 0.3$. An escape fraction $>1$ is unphysical because it implies the galaxy is leaking more than the ionizing photons produced in this ionization cone. Note that the properties of the stellar population might not be the same throughout the galaxy. Even assuming a more typically used $L_{1500}/L_{900} = 3$, we get $f_{\text{esc}}^{\text{rel}} = 1.2\pm 0.3$ and $f_{\text{esc}}^{\text{abs}} = 0.6\pm 0.1$. Using the lower limit on $L_{1500}/L_{900}= 1.25$ gives us the most conservative estimate on the relative escape fraction of $f_{\text{esc}}^{\text{rel}} = 0.5\pm 0.1$ and a corresponding $f_{\text{esc}}^{\text{abs}} = 0.24\pm 0.06$. A future paper will analyze the resolved stellar population properties of z19863 and test the impact different stellar population models have on the escape fraction measurements.

\begin{figure*}
	\centering
	\tiny
	\includegraphics[scale=0.65, trim=0.0cm 0.0cm 0.0cm 0.0cm,clip=true]{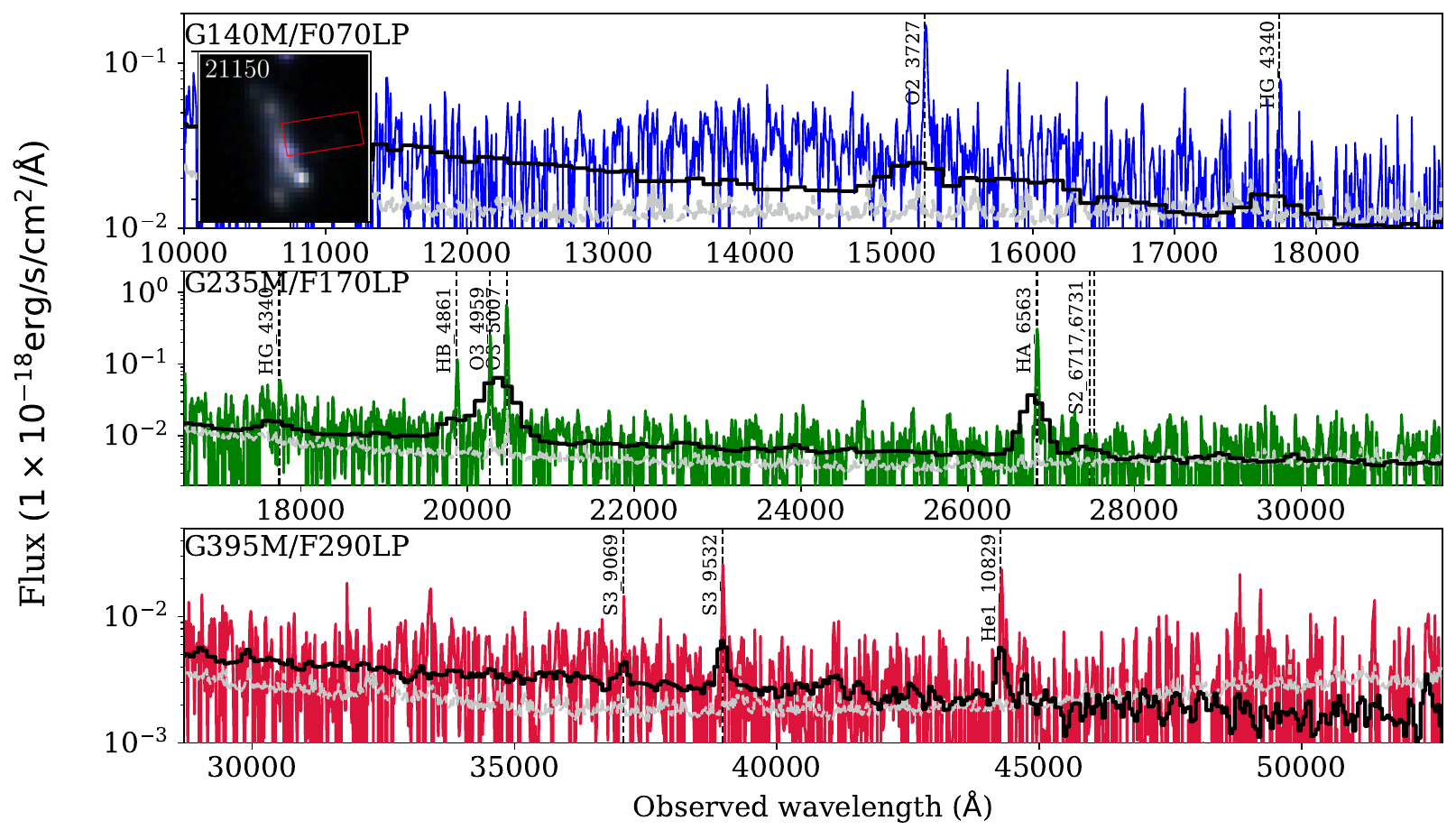}		
\caption{NIRSpec spectrum (JADES NIRSpec ID = 21150, NIRCam ID = 211968) from the JADES in the three medium band grisms, G140M/F070LP (top), G235M/F170LP (middle), and G395M/F290LP (bottom). The grey dashed curve in each panel corresponds to the error spectrum in the respective grism.  The low-dispersion prism spectrum is shown as the solid black curve in each panel and has significantly lower error spectrum than displayed ranges here. The black dashed lines mark the location of bright emission lines detected above the $5\sigma$ level in each sub-panel. The inset image in the top panel shows the location of the slit with respect to the JWST/NIRCam image \citep[taken from the Figure B.1 in ][]{bunker_2023_JADESNIRSpecInitial}.
}
	\label{fig:nirspec_spec}
\end{figure*}

\subsection{Merger activity}\label{sec:merger}

We use medium-band filters F210M and F182M to analyze the merger activity of our source. At $z=3.088$, the medium-band filter F210M would contain emission from stars and the \oiii+\hb\ emission lines, whereas filter F182M would only have stellar emission. Therefore, the difference image between the two filters provides a spatially resolved distribution of gas emission (\oiii+\hb) from the target (Figure \ref{fig:fresco_pseudo_o3}). We use StatMorph \citep{rodriguez-gomez_2019_OpticalMorphologiesGalaxies} to estimate non-parametric quantities for all three filters (more details in Gupta et al., in prep.) .

The Gini-M20 statistics \citep{lotz_2008_EvolutionGalaxyMergers} classify z19863 as a merger based on the stellar emission (F182M) but a non-merger based on the gas emission (F210M-F182M). The peak gas emission (F210M-F182M) is coincident with the peak non-ionizing UV emission (F606W), whereas the peak stellar emission (F182M) is $0.18''$ SW of the non-ionizing UV emission (F606W). The lack of merger signs in the gas emission could be because hydrodynamical cooling results in faster settling time for the gas compared to stars after a merger event (See Gupta et al., in prep. for more discussion). The gas emission (F210M-F182M) seems to be extended towards the position of the LyC detection, indicating that LyC photons could be escaping from a recently star-forming part of the galaxy. A detailed analysis of spatially-resolved spectral properties of z19863 is beyond the scope of this paper and will be part of future analysis

\begin{figure*}
	\centering
	\tiny
 	\includegraphics[scale=0.38, trim=0.0cm 0.0cm 0.0cm 0.0cm,clip=true]{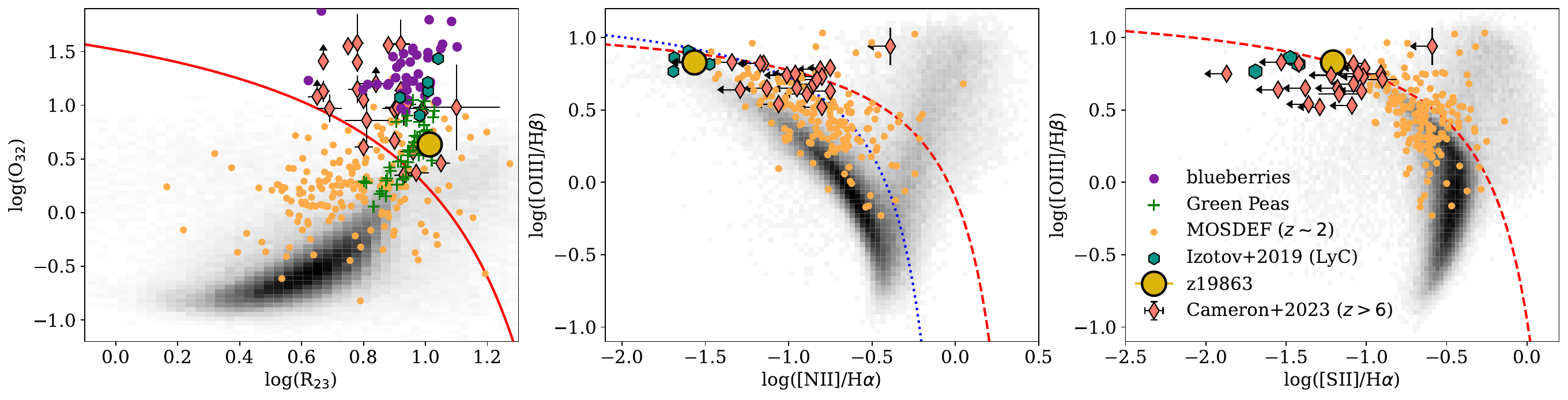}
\caption{The location of z19863 (golden circle) on the $R23-O32$ (left) and $N2-R2$ (middle) and $N2-R2$ (right) diagrams, typically is used to probe the metallicity and ionization conditions of the ISM. The $z\sim 6$ galaxies from \cite{cameron_2023_JADESProbingInterstellar} are shown as salmon diamonds.  We also compare with other extreme galaxies in the local universe (blueberries from \cite{yang_2017_BlueberryGalaxiesLowest}: purple circles, green peas from \cite{yang_2017_LyaProfileDust}: green plus, and LyC leakers from \cite{izotov_2018_LowredshiftLymanContinuum}: green hexagons). The grey shaded regions and the orange dots in each panel represent the typical SFGs at $z\sim0$ from the SDSS survey  \citep{aihara_2011_EighthDataRelease} and at $z\sim2$ from the MOSDEF survey \citep{kriek_2015_MOSFIREDEEPEVOLUTION} respectively. The solid red curve in the left panel and dashed red curve in the middle and right panels shows the \cite{lamareille_2004_LuminositymetallicityRelationLocal} and \citep{kewley_2001_TheoreticalModelingStarburst} relation for identifying AGNs respectively. The blue dotted curves in the middle panel represents the  \cite{kauffmann_2003_HostGalaxiesAGN} relation for star-forming galaxies. Our galaxy falls in the star-forming region occupied by metal-poor galaxies, similar to $z>6$ galaxies and other extreme galaxies in the local universe. }
	\label{fig:bpt}
\end{figure*}

\subsection{JADES/NIRSpec observations}\label{sec:nirspec}
Figure \ref{fig:nirspec_spec} shows the JWST/NIRSpec spectrum for z19863 from the \jades\ \citep[NIRSpec ID = 21150, NIRCam ID = 211968;][]{bunker_2023_JADESNIRSpecInitial} spanning all the way from $1-5\mu m$, showing multiple bright emission lines. Note that the $0.46''$ microshutter only partially covers the galaxy and only on the side opposite to the LyC detection. We cannot use the NIRSpec spectra to rule out if the emission in the F336W image is due to an interloper galaxy. Even with this partial coverage, we are able to detect a whole suite of emission lines to estimate the gas-phase metallicity and ionization conditions in the galaxy.

The emission lines fluxes in the \jades\ catalogs have been aperture-corrected using the location of micro-shutter on the NIRCam imaging. The analysis in the following subsections assumes that the gas has similar properties across the entire galaxy, which might not be correct. Neither \nii\ or \sii\ are detected at S/N$>5$ with grism, therefore we rely on the \jades\ prism catalogs to estimate ISM conditions. 

\subsubsection{Dust Extinction and star formation}\label{sec:dust}

We use the \cite{gordon_2003_QuantitativeComparisonSmall} extinction curve for the Small Magellanic Cloud to correct for dust extinction, following similar approach used in previous studies of high redshift galaxies \citep{reddy_2023_PaschenlineConstraintsDusta, cameron_2023_JADESProbingInterstellar}. 
 We use the Balmer decrement based on the ratio of \halpha\ and \hb\ emission lines to estimate the color excess using the equation:
 \begin{equation}
    E(B-V) = \frac{2.5}{\kappa(\hb) - \kappa(\halpha)}\log_{10}\left(\frac{(\halpha/\hb)_{\rm obs}}{2.98}\right),
\end{equation}
The Balmer decrement for z19863 is $f(\halpha/\hb)_{\rm obs} = 3.26\pm0.07$, which results in $E(B-V) = 0.09\pm0.02$. 

The dust-corrected SFR based on \halpha\ and using the \cite{kennicuttjr_2012_StarFormationMilky} relation, is $4.3\pm 0.09$ \msun/yr for this galaxy, placing it within the main-sequence at $z=3$ \citep{popesso_2022_MainSequenceStarforming}. It's important to note that the \halpha\ luminosity used for this calculation is rescaled based on the position of the microshutter on the NIRCam image of the galaxy. However, the pseudo \oiii\ emission map reveals that the microshutter misses the most star-forming part of the galaxy (Figure \ref{fig:fresco_pseudo_o3}). Therefore, the total \halpha\ luminosity and the instantaneous SFR of z19863 may differ significantly from the value estimated here.

\subsubsection{ionization conditions}

The left panel in Figure \ref{fig:bpt} displays the extinction-corrected $R23$ versus $O32$ relation of z19863 compared to other LyC emitter samples and galaxies at $z>6$, where:
$$R23= \log_{10}((\oiii,4959+\oii)/\hb),$$ 
$$O32 =  \log_{10}(\oiii/\oii).$$
To correct for dust extinction for low redshift samples, we apply the CCM89 extinction curve \citep[$R_V=3.1$, ][]{cardelli_1989_RelationshipInfraredOptical}. The $O32$ and $R23$ ratios of z19863 are higher than those of typical star-forming galaxies both in the local universe \citep{aihara_2011_EighthDataRelease} and at $z\sim 2$ \citep{shapley_2014_MOSDEFSurveyExcitation, 
kriek_2015_MOSFIREDEEPEVOLUTION}. The emission line ratios of z19863 resemble those of other extreme galaxy populations such as green peas and blueberries in the local universe. However, its $O32$ ratio is not as high as that of other LyC emitters at $z\sim0.3$. z19863 also exhibits a low $S2$, where $$S2 = \log_{10}((\sii)/\halpha),$$ compared to typical star-forming galaxies at $z\sim 2$, which has also been indicated as a marker for LyC leakers \citep{ramambason_2020_ReconcilingEscapeFractions}.
 

Interestingly, we have detections of both \siii\ and \sii\ in the prism spectrum of z19863. Therefore, we can use the $S32$ ratio defined as:
$$S32 = log_{10}(\siii/\sii),$$
to estimate the ionization parameter, a metallicity- and pressure-independent ionization parameter diagnostic. Traditionally, the $O32$ ratio is used to estimate the ionization parameter because \siii\ either redshifts into the near-infrared at $z>0.05$ or beyond $2.4\mu m$ at $z>1.5$, making it difficult to observe. However, ionization parameters based on $O32$ are highly dependent on the choice of stellar template \citep{sanders_2015_MOSDEFSurveyElectron}. \cite{sanders_2019_MOSDEFSurveyIii} used stacked measurements of \siii\ in $z\sim 1.5$ galaxies to and confirmed its weaker dependence on metallicity compared to $O32$.
 Recent observations have started to use \siii\ to estimate the ionization conditions in galaxies \citep{mingozzi_2020_SDSSIVMaNGA, kumari_2021_HardnessIonizingRadiation}.

 The $S32$ ratio of z19863 is $0.27\pm0.09$ constraining the ionization parameter to $logU=-3.25\pm0.10$. This suggests that z19863 has a similar ionization parameter to typical SFGs at $z\sim2-3$ \citep{sanders_2015_MOSDEFSurveyElectron}, and significantly lower than galaxies at $z>6$. Again, we note that NIRSpec observations only cover a small portion of the galaxy. Spatially resolved observations of star-bursting galaxies find significant variations in metallicity and ionization parameter \citep{delvalle-espinosa_2023_SpatiallyResolvedChemodynamics}, indicating that NIRSpec spectra might not be representative of the average gas conditions in the galaxy.


\subsubsection{Gas-phase Metallicity}
The \nii\ emission line is not detected in either the grism or prism spectrum, indicative of very low metallicity. We use the grism spectra to estimate an upper limit on \nii\ because \nii\ is blended with \halpha\ in the prism spectrum. The \nii/\halpha\ versus \oiii/\hb\ diagram (Figure \ref{fig:bpt}) places the galaxy well within the region occupied by low-metallicity star-forming galaxies both in the low and high-redshift universe.

We use the S23 diagnostic from \citep{kewley_2019_TheoreticalISMPressure} to estimate the gas-phase metallicity, where 
$$S23 = log_{10}((\sii+\siii)/\halpha),$$
due to its sensitivity at lower metallicity and the lack of \nii\ detection.  Using an $S23$ ratio of $-0.75\pm0.05$ and $\log U=-3.25$, we estimate the gas-phase metallicity at $12+\log(O/H)=7.79_{-0.05}^{+0.06}$ ($Z= 0.13\pm0.01$\,Z$_{\odot}$). Thus, the gas-phase metallicity of our galaxy is fairly consistent with gas-poor galaxies detected at higher redshifts \citep{tang_2023_JWSTNIRSpecSpectroscopy, boyett_2024_ExtremeEmissionLine}.      

\begin{figure*}
	\centering
	\tiny
	\includegraphics[scale=0.55, trim=0.0cm 0.0cm 0.0cm 0.0cm,clip=true]{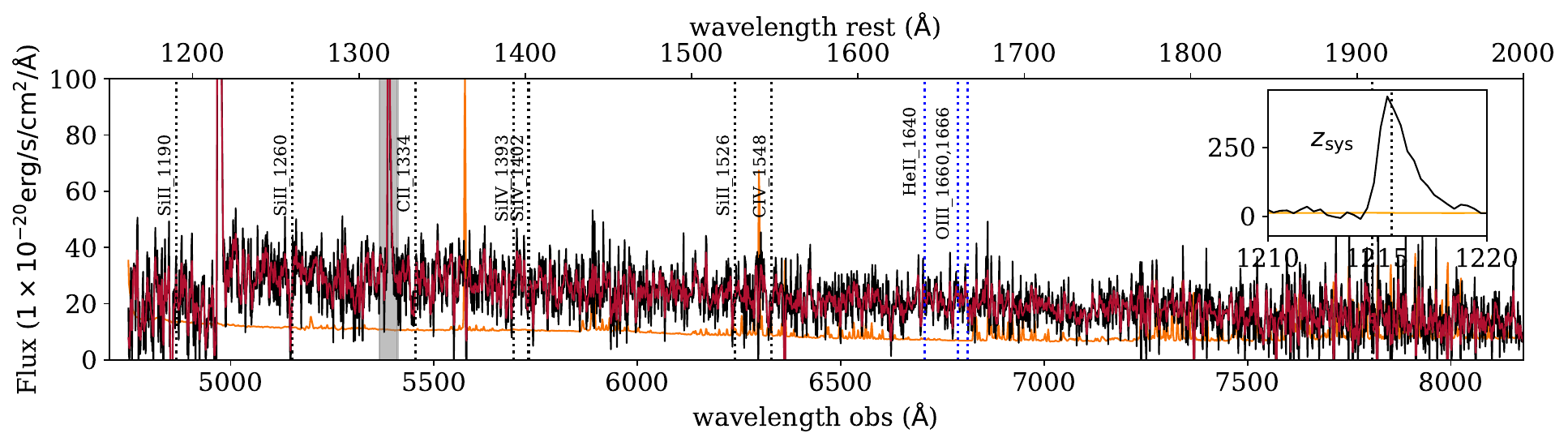}		
\caption{Rest-UV 1D spectrum extracted from the MXDF mosaic (12 hours integration). The inset image shows the zoomed-in spectrum at the location of \lya\ emission line. The dashed vertical lines mark the location of various absorption features. The dashed orange curve is the corresponding error spectrum. The grey shaded region highlights the \lya\ emission from the nearby $z=3.4$ galaxy.}	
\label{fig:muse_spec}
\end{figure*}

\subsection{Rest-UV spectrum}\label{sec:muse}
We extract the rest-UV spectrum of our target using the mosaic datacube from the MXDF survey \citep{bacon_2023_MUSEHubbleUltra}.  The mosaic datacube reaches a total integration time of 12 hours at the position of our target (peak flux in F606W filter). We search for emission around the expected \lya\ wavelength at $z=3.088$, and identify the peak wavelength.   We sum over flux within $\pm 10\,\Angstrom$ around the peak \lya\ wavelength to create a line map and add spaxels above $2\times\ {\rm RMS_{sky}}$ to extract the 1D spectrum (Figure \ref{fig:muse_spec}). The peak \lya\ emission has a spatial offset of $0.''42$ with respect to peak flux in F606W filter, which is similar to the typical FWHM of the MUSE observations.

The \lya\ emission is highly asymmetric with an undetected blue-side, similar to other high-redshift \lya\ emitters \citep{gronke_2016_LymanAlphaSpectraMultiphase}. The red peak of the \lya\ emission has a marginal velocity offset from the systematic redshift of the galaxy (approximately 100 km/s). There is significant contamination in the rest-UV spectrum from another \lya\ emitter at $z=3.4$, which is only $0.85''$ from the F606W peak (indicated as a black plus in Figure \ref{fig:fresco_imaging}). The high spatial resolution of HST and JWST ensures negligible contamination in the photometry of z19863 from this source at any wavelengths including the F336W image. We have clear detections of many low (\SiIIAbsA, \SiIIAbsB, \CIIAbs, \SiIIAbsC) and high ionization (\SiIVAbs, \CIVAbs) metal absorption features. Interestingly, we do not detect any high ionization emission lines such as \oiiiuv, \heii, \ciii, which are commonly detected in other LyC emitters \citep{vanzella_2019_IonisingIntergalacticMedium}. \cite{ji_2020_HSTImagingIonizing} also report a tentative detection of \heii\ (S/N$\sim 2$) in Ion1, indicating the need for a deeper rest-UV spectrum.

\section{Discussion}\label{sec:discussion}

\subsection{Curious case of z19863}
This paper presents the discovery of a new LyC emitter candidate, z19863 at redshift $z=3.088$, combining archival data from deep imaging campaigns conducted on the HST and JWST (Figure \ref{fig:fresco_imaging}). Detailed information about the source is provided in Table \ref{tb1}. Using the deep NIRSpec spectrum from \jades\, we find that z19863 has characteristics similar to other LyC emitters across different redshift ranges (see Figure \ref{fig:bpt}), including a moderate ionization parameter ($\log U = -3.25 \pm 0.1$), a low gas-phase metallicity (12+$\log$(O/H) = $7.76 \pm 0.06$), and a high $O32$ ratio ($3.65 \pm 0.22$). Notably, the LyC emission is spatially offset by $0.29'' \pm 0.04''$ (equivalent to $7 \pm 1$ pixels or approximately 2.2 kpc) from the peak rest-UV (non-ionizing) emission (Figure \ref{fig:fresco_imaging}). The astrometric misalignment is insufficient to account for the spatial offset ($dx = \pm 1$ pixel). We do not detect any interloper galaxy co-incident with the LyC detection in extremely deep spectroscopy from the  MXDF survey (Figure \ref{fig:muse_spec}). We estimate the probability of randomly detecting a similar source within a similar aperture ($0.24''$) in the F336W image is  $p<6\times 10^{-5}$.

Most observations focused on detecting LyC photons typically search for emission at $\lambda_{rest} < 912\,\Angstrom$ co-spatial with the non-ionising UV emission. For example, \cite{kerutt_2023_LymanContinuumLeaker} look for LyC emission within $0.''35$ aperture around the non-ionising rest-UV emission, and therefore would miss a source similar to z19863 with about $0.''29$ offset between the ionising and non-ionising emission peaks. Ion1 ($z=3.7$) is another know LyC emitter that exhibits a spatial offset between the peak rest-UV and LyC emission \citep[$0.''12 \sim 0.85$\,kpc, ][]{ji_2020_HSTImagingIonizing}. 


Additionally, z19863 has a single peaked \lya\ emission profile z19863 (Figure \ref{fig:muse_spec}), atypical to LyC emitters in the local universe where a double-peaked profile with a narrow velocity separation is common \citep{izotov_2018_LowredshiftLymanContinuum, izotov_2022_LymanAlphaLyman, flury_2022_LowredshiftLymanContinuum}. Models of LyC and \lya\ escape suggest that a narrow velocity separation is indicative of an optically thin neutral ISM, which, in turn, facilitates the escape of ionising photons \citep{izotov_2018_LowredshiftLymanContinuum}.  The green pea galaxies with high $O32$ ratios (are more likely to leak LyC photons) have lower HI gas mass, suggesting they are more likely to be optically thin \citep{kanekar_2021_AtomicGasMass}. 
However, LyC emitters at high redshift exhibit a range of \lya\ profiles, including 3-4 peaks \citep{izotov_2022_LymanAlphaLyman}, single peaked for Q1549-C25 \citep{shapley_2016_Q1549C25CleanSource}, and absorption rather than emission for Ion1 \citep{ji_2020_HSTImagingIonizing}. \cite{kerutt_2023_LymanContinuumLeaker} uses a larger sample of LyC emitters between $3<z<4$ and find no correlation between the \lya\ emission and the escape fraction. It is possible that at higher redshifts, most ionising photons escape because of the inhomogeneities in the ISM rather than an optically thin ISM.



\subsection{Mergers driving the LyC escape}

We currently have two models explaining LyC escape: through an optically thin ISM and the picket-fence model \citep{zackrisson_2013_SpectralEvolutionFirst}. In the picket-fence model, LyC photons escape through a few optically thin channels in an otherwise optically thick ISM. These channels are typically created by star-formation-triggered outflows capable of blowing out neutral gas. LyC emitters in the local universe often show a significant correlation between the escape fraction and the strength of ionized gas outflows, suggesting that star formation feedback is responsible for creating these ionized gas channels \citep{amorin_2024_UbiquitousBroadlineEmission}.

However, the mass loading factors for most high-redshift galaxies are similar to those of local dwarf galaxies, even if their star formation rates (SFRs) are ten times higher \citep{gupta_2022_MOSELSurveyExtremely, concas_2022_BeingKLEVERCosmic, llerena_2023_IonizedGasKinematics, carniani_2023_JADESIncidenceRate}. Some semi-analytic models of star-formation-driven galactic winds suggest that outflowing gas in super star-forming conditions might start radiatively cooling on a relatively short timescale \citep[$<1\,$Gyr, ][]{lochhaas_2018_FastWindsDrive}. This rapid cooling might hinder the creation of large-scale channels of optically thin neutral gas solely from star formation alone.

Mergers and galaxy-galaxy interactions are known to tidally strip gas and stars, leading to the formation of highly inhomogeneous galaxies \citep{toomre_1972_GalacticBridgesTails, cox_2008_EffectGalaxyMass, spilker_2022_StarFormationSuppresion}. Very few LyC emitters in the local universe are close enough to spatially map their neutral gas density. Deep 21cm observations on Haro 11, a local LyC emitter, revealed pockets of optically thin neutral hydrogen most likely created by a recent merger event \citep{reste_2023_TidallyOffsetNeutral}. The large spatial offset (6 kpc) between the peak neutral gas density and the LyC emitting clump suggests that LyC photons are most likely leaking through these optically thin channels \citep{reste_2023_TidallyOffsetNeutral}. On the other hand, two other local LyC leakers with 21cm observations exhibit either only an upper limit on HI mass \citep{puschnig_2017_LymanContinuumEscape} or very high gas mass \citep{haynes_2018_AreciboLegacyFast}, indicating a wide range in the properties LyC emitters in the local universe. 


Recent work by \cite{gupta_2023_MOSELSurveyJWSTa} finds that EELGs at $z\sim 3.5$, analogs of galaxies at $z>6$, are more likely to encounter major mergers and/or strong interactions. Other observations \cite[e.g.,][]{duncan_2019_ObservationalConstraintsMerger} also find increased merger activity at higher redshifts. For z19863, medium band images from the FRESCO survey indicate that the galaxy might have encountered a merger recently (Figure \ref{fig:fresco_pseudo_o3}). \cite{bassett_2018_LackCorrelationOIII} suggest that the lack of correlation between the O32 ratio and escape fraction could be due to anisotropic neutral gas coverage. In fact, all of their LyC candidates with the lowest O32 ratios showed evidence of disturbed morphology. The \lya\ emission peak for the candidate LyC emitters in \cite{kerutt_2023_LymanContinuumLeaker} is spatially offset by $0.3''-0.5''$ from the peak LyC emission, indicating a mismatch between the peak neutral gas density and ionising stars. We suspect that inhomogeneities created by a merger event in the neutral ISM might have played a significant role in aiding LyC photons to escape at higher redshifts.

We think in z19863 ISM is neither optically thin everywhere nor has a complete picket-fence structure with many optically thin ionizations cone. Here the ISM most likely has a single optically thin ionization cone of neutral ISM created by a recent merger event. The ionising photons are leaking through this single channel, and by happenstance, its alignment along our line of sight ensures that we can detect the faint signature of ionising radiation leaking from this single ionization cone. 


\begin{figure*}
	\centering
	\tiny
 	\includegraphics[scale=0.55, trim=0.0cm 0.0cm 0.0cm 0.0cm,clip=true]{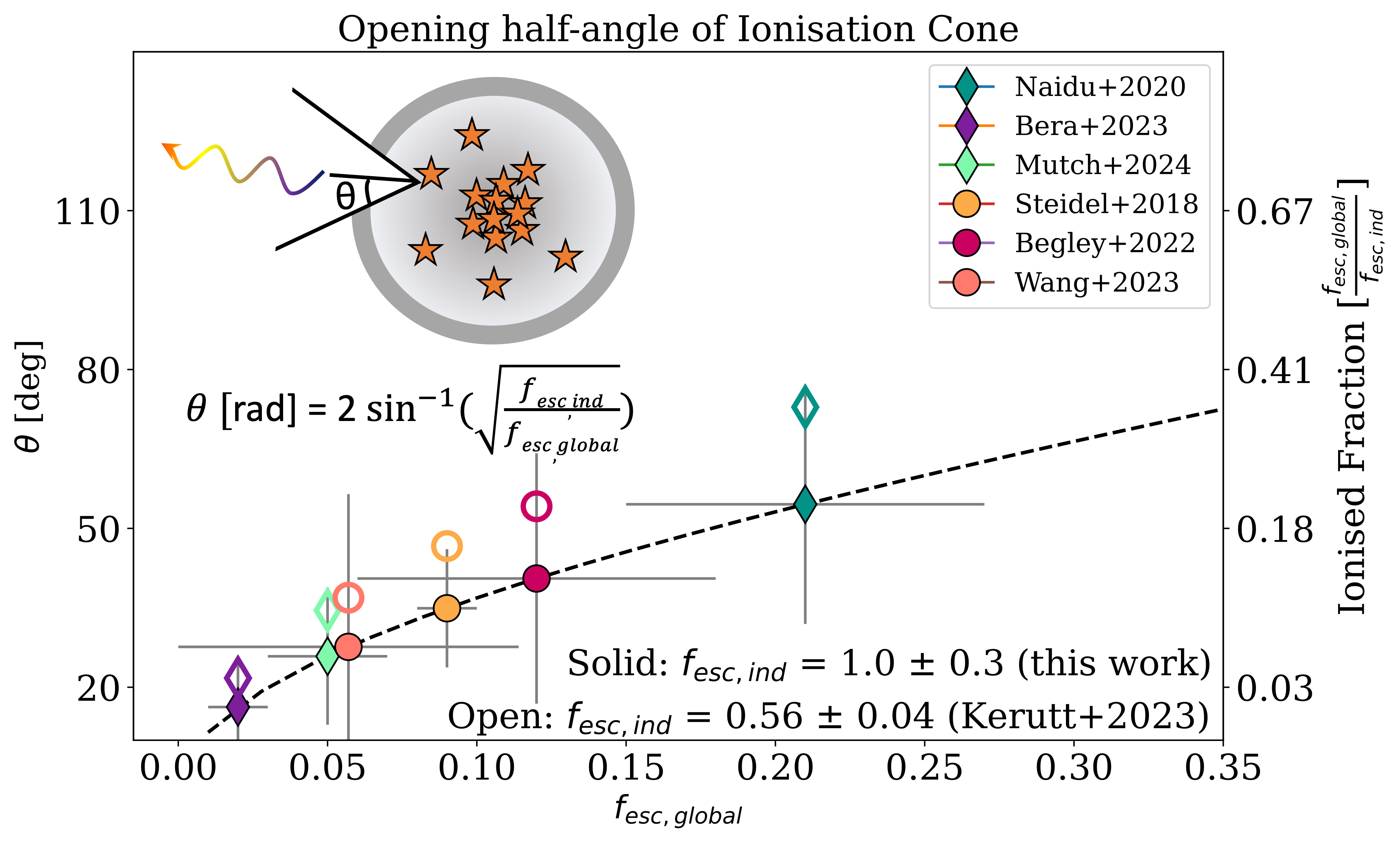}

\caption{Relation between the opening half-angle of the ionization cone and the global \fesc\ values. The diamonds (purple: \cite{bera_2023_BridgingGapCosmic}, light green: \cite{mutch_2023_DarkagesReionizationGalaxy}, and  dark green \cite{naidu_2020_RapidReionizationOligarchs}) and circles (salmon: \cite{wang_2023_LymanContinuumEscape}, yellow: \cite{steidel_2018_KeckLymanContinuum}, and magenta: \cite{begley_2022_VANDELSSurveyMeasurementa}) show the theoretical  and observational limits on the global \fesc\ respectively. The solid and open symbols use individual \fesc\ based on z19832 (this work) and average \fesc\ from five gold LyC candidates from \cite{kerutt_2023_LymanContinuumLeaker}. The black dashed line is for the fully ionised channels ($f_{esc, ind} =1$). }	
\label{fig:opening_angle}
\end{figure*}

\subsection{Modelling the opening angle of ionising channels}

If LyC radiation leak through a few optically thin ionization cones, similar to z19832, for most galaxies, then their random alignment along the line of sight will dictate the number of confirmed detections of LyC emitters. Estimating the total covering fraction of optically thin ISM in high-redshift galaxies is extremely challenging. \lya\ photons get fluorescently scattered by neutral hydrogen, and thus, a \lya\ emission profile provides some hints about the neutral gas density. A small velocity separation and a large ratio of total flux between the blue and red peaks usually indicate an optically thin ISM \citep{erb_2018_KinematicsExtendedLymanalpha, blaizot_2023_SimulatingDiversityShapes, mukherjee_2023_CompactExtendedLyman}. However, spatially resolving \lya\ emission is only possible for a small number of \lya\ emitters at $z>3$, until future spectrographs such as MAVIS \citep{mcdermid_2020_PhaseScienceCase} become available.

Stacking experiments with large samples of SFGs at $z\sim3$ have provided some constraints on the average \fesc$=1-10\%$  \citep[][]{steidel_2018_KeckLymanContinuum, wang_2023_LymanContinuumEscape}.   Theoretical models using various observations about the reionization history of the universe, such as neutral gas density, UV luminosity function etc.,  constrain the average \fesc\ required \citep{naidu_2020_RapidReionizationOligarchs, bera_2023_BridgingGapCosmic, mutch_2023_DarkagesReionizationGalaxy}. However, some individual LyC emitters can have escape fraction as high as 100\% \citep{rivera-thorsen_2022_BottomupSearchLymancontinuum, kerutt_2023_LymanContinuumLeaker}.

We can use individual \fesc\ measurements to develop a simple model for the total covering fraction or the opening angle of the ionization cone required to reach global \fesc\ values.  We assume that all galaxies in the universe have on average an ionization cone with solid angle  $\Omega$ and escape fraction $f_{esc, ind}$. The following equation balances the total ionization flux from a single ionization cone with the average escape fraction expected over the entire galaxy ($f_{esc, global}$)
\begin{equation}
    \Omega \times f_{esc, ind} = 4\pi \times f_{esc, global}, 
    \implies \Omega = 4\pi \times \frac{f_{esc, global}}{f_{esc, ind}}. 
\end{equation}
For a cone with apex half angle of $\theta$, the total solid angle is 
\begin{equation}
\Omega = 4\pi \sin^2{\frac{\theta}{2}}.
\end{equation}
Therefore, the half-angle of ionization cone would be 
\begin{equation}\label{eq:theta}
\theta [rad] = 2\times \sin^{-1}{\sqrt{\frac{f_{esc, global}}{f_{esc, ind}}}}. 
\end{equation}

Figure \ref{fig:opening_angle} shows the average half-angle estimated using equation \ref{eq:theta} based on $f_{esc,ind}$ of z19863 and various measurements of $f_{\text{esc, global}}$ from observations \citep{steidel_2018_KeckLymanContinuum, wang_2023_LymanContinuumEscape} and theory \citep{naidu_2020_RapidReionizationOligarchs, bera_2023_BridgingGapCosmic, mutch_2023_DarkagesReionizationGalaxy}. By definition the escape fraction cannot be greater than 1, therefore we assume an absolute escape fraction of $1.00\pm 0.3$ rather than estimated value of $f_{esc}^{abs} = 1.25\pm 0.3$ in Section \ref{sec:lyc}. Additionally, the opening half-angle is estimated using $f_{\text{esc, ind}}$ based on the average escape fraction for the five gold LyC emitter candidates from \cite{kerutt_2023_LymanContinuumLeaker}.


Our simple model can help distinguish between more likely global escape fraction measurements.  For instance, if the global escape fraction is constrained to be very low, such as $0.02_{-0.01}^{+0.04}$ \citep{bera_2023_BridgingGapCosmic}, an ionization cone with an opening half-angle of approximately $\sim16^{\circ}$ may be sufficient to account for this small escape fraction. Even if only 50\% of ionizing photons leak from the ionization cone of most galaxies, the opening half-angle would only need to increase to approximately $23^{\circ}$. This scenario could provide an explanation for the limited number of LyC emitter detections at higher redshifts.
On the other hand, to achieve a global escape fraction of $0.2$ \citep{naidu_2020_RapidReionizationOligarchs}, ionization cones with $f_{\text{esc, ind}} = 1.0$ [$f_{\text{esc, ind}} = 0.5$] will need to have opening half-angle of $\sim 53^{\circ}$ [$78^{\circ}$]. Such large opening angles imply that a significant fraction (20-50\%) of the galaxy would need to be leaking ionizing radiations. However, if this were the case, one might expect a larger success rate in building a larger sample of confirmed LyC leakers. 

We only need an opening half-angle of $\sim37^{\circ}$ for fully ionised cone and $\sim 53^{\circ}$ for ionization cone with 50\% escape fraction to explain the typically observed \fesc\ from observations \citep{steidel_2018_KeckLymanContinuum, wang_2023_LymanContinuumEscape}. It's worth noting that even at $z=3$, a partially neutral intergalactic medium (IGM) can absorb a considerable fraction of ionizing radiations \citep{bassett_2021_IGMTransmissionBias}, further complicating the detection of LyC emission. Altering the median opening angle of escaping ionising photons would have an impact on the topology of reionization, and a more sophisticated model would explore the effects of this change on the progress of reionization.


\section{Summary}

This paper presents discovery of a new candidate LyC emitter (z19863) at $z=3.088$, where LyC emission is detected using the F336W/UVIS filter on the HST. The galaxy is part of the EELG sample in the MOSEL survey \citep{tran_2020_MOSELStrongOiii, gupta_2022_MOSELSurveyExtremely}. The LyC emission is spatially offset by $0.''29\pm0.''04\ \sim\ 2.2\pm0.3\,$kpc from the peak rest-UV emission (F606W). Based on the NIRSpec spectra from the JADES, we estimate that the galaxy has low metallicity, high ionization parameter, and low dust attenuation (See Table \ref{tb1}), similar to other LyC emitters at both low and high redshifts. 

The deep rest-UV spectrum of the galaxy from the MXDF survey (12 hour integration) exhibit a single peaked \lya\ emission profile, and a range of low- and high- ionization absorption features (Figure \ref{fig:muse_spec}). The lack of the rest-UV emission lines such as \oiiiuv\, \civ, \ciii\ etc. that are typically used as proxy for LyC emission \citep{schaerer_2022_StrongLymanContinuum}, suggest an ISM conditions in this galaxy are dissimilar to a typical LyC emitter in the local universe. Further analyses of the absorption profiles of various species and their correlation with LyC emission \citep{mauerhofer_2021_UVAbsorptionLines} is beyond the scope of this paper and will be part of future work.

We propose that LyC photons in z19863 are leaking through a narrow cone of optically thin ISM, likely created by a recent merger event as indicated by the disturbed morphology in the medium band imaging (see section \ref{sec:merger}). The ionization cone is spatially offset by $\sim0.''3$ from the non-ionising UV emission and its chance alignment along the line of sight ensures its detection by us. A source similar to z19863 would be missed by studies looking for LyC emission co-incident with peak UV emission \citep[e.g.,][]{kerutt_2023_LymanContinuumLeaker}. Using a simple toy model, we estimate that the opening half-angle of ionization cones can be as low as $16^{\circ}$ to explain some of the global measurements of escape fraction, indicating that very small fraction of the galaxy needs to be leaking ionising radiation. Such small opening angle would also explain the relatively low number of confirmed LyC emitters at high redshifts.

\section{Acknowledgments}

This research were supported by the Australian Research Council Centre of Excellence for All Sky Astrophysics in 3 Dimensions (ASTRO 3D), through project number CE170100013. 

%

\vspace{5mm}









\end{document}